\documentstyle[12pt]{article}
\begin{document}
\title{Coupled-Channel Final-State Interactions through Reggeon Exchange 
for $D(B) \rightarrow \pi \pi , K\bar{K}$}
\author{
{P. \.Zenczykowski}$^*$\\
\\
{\em Dept. of Theoretical Physics} \\
{\em Institute of Nuclear Physics}\\
{\em Radzikowskiego 152,
31-342 Krak\'ow, Poland}\\
}
\maketitle
\begin{abstract}
Coupled-channel final-state-interaction effects for $D$ and $B$ weak
decays into $\pi \pi$ and $K\bar{K}$ 
are discussed in a Regge framework.
It is found that the inclusion of coupled-channel effects significantly affects
the results obtained previously in a quasi-elastic approximation.
It is also shown that in the isospin $I=0$ channel
the inelastic final-state transitions
$(\pi \pi)_{I=0} \rightarrow (K\bar{K})_{I=0}$ 
dramatically influence the phase of
the $B^0 \rightarrow (K\bar{K})_{I=0}$ amplitude.

\end{abstract}
$^*$ E-mail:
zenczyko@solaris.ifj.edu.pl
\newpage

Gathering data on CP-violation through detailed studies of nonleptonic 
decays of B mesons becomes one of the principal goals in our attempt to unravel 
the mystery of CP-violation.
It is generally hoped that experimental results on nonleptonic B decays
will provide information that will allow us to decide
whether the standard model is correct or not.
Unfortunately, extracting fundamental CP-violating parameters 
of the standard model from various B decay modes is not trivial at all.
One of the main problems encountered 
is a reliable estimate of hadronic effects in the final state.
Although such final-state-interaction (FSI) effects are often considered
to be of no particular 
importance by themselves, their determination is crucial 
for the success
of the whole project of extraction of fundamental CP-violating parameters.

In recent years several authors have stressed the importance of FSI in B decays
\cite{Don86}-\cite{Neubert98}.
%\cite{Don86,Wolf95,Don96etal,Zen97,Neubert98}.  
Although there is a growing
understanding that FSI must be taken into account, there are severe problems
with their reliable treatment.  Various approaches have been considered 
\cite{VarFSI}-\cite{GRrescatt}.  
%\cite{VarFSI,CSmith,GRrescatt}.  
Usually only some intermediate states, believed to provide the 
largest effects, are taken into account.  A Regge-model-based approach is often
used here to estimate high-energy interactions between decay products.
Recently, a simple model of this type has been applied to the description
of strong phases in $D(B) \rightarrow K \pi$ and $D(B) \rightarrow \pi \pi $, $
K \bar{K}$ decays \cite{GW97,DGPW98,GPW98}.

The model of refs.\cite{GW97,DGPW98,GPW98} 
is based on a quasi-elastic approximation.
This approximation considers rescatterings of the type:
$K \pi \rightarrow K \pi$, 
$\pi \pi \rightarrow \pi \pi$, $K \bar{K} \rightarrow K \bar{K}$.
All other possible final-state interactions are ignored.
The model yields strong-interaction phases which compare favourably with
the numbers extracted (with the help of some simplifying assumptions) 
from the existing experimental data at $D$ mass
\cite{CLEO}, and predicts these phases at
$B$ mass.
When assessing the reliability of these predictions, one may of course question 
the assumption of restricting the intermediate states to those composed of
two pseudoscalar mesons only. However, even when one accepts that the
contribution from these states dominates, or that the contribution from other  
states simply follows the pattern of the $PP$ contribution ($P$ denotes a
pseudoscalar meson), 
the question still remains whether the
inelastic FSIs, ignored in \cite{DGPW98,GPW98}, modify their predictions
in an essential way.
Inelastic FSI means here the coupled-channel effects of the type
$\pi \pi \rightarrow K \bar{K}$ or $K \pi \rightarrow K \eta $.
Inclusion of such processes has been shown to be very important if a fully 
SU(3)-symmetric FSI-including effective-quark-diagram 
description of weak nonleptonic decays is to be achieved
\cite{Zen97} (see also ref.\cite{GRrescatt}).
In this paper we analyse in some detail the influence of coupled-channel
effects on the predictions of the quasi-elastic Regge approach of ref.
\cite{DGPW98,GPW98}. 
In papers \cite{DGPW98,GPW98} SU(3) symmetry was broken.
Since our whole coupled-channel
approach is calculationally simple 
only when SU(3) symmetry is unbroken, 
we will keep SU(3) symmetry throughout this paper.
We will show, however, that in 
the no-coupled-channels case the original and SU(3)-symmetric versions 
of the model of ref.\cite{GPW98} do not differ much in their predictions.
We will restrict ourselves to the noncharmed, nonstrange
sector of two-meson interactions, 
i.e. to the analysis of coupled-channel effects in the $\pi \pi$, 
$K \bar{K}$, $\pi ^0 \eta _8$
and $\eta _8 \eta _8$ channels.
The latter two channels are
included both because they are needed to maintain SU(3) symmetry of the
analysis, and because the effects of these channels should be comparable to that
of the $K \bar{K}$ channel 
(the mass of $\eta \approx \eta _8$ is close to that of the kaon).  

Calculations within the Regge approach of refs. \cite{DGPW98,GPW98} take into
account the Pomeron and the exchange-degenerate Reggeons $\rho $, $f_2$, 
$\omega$, and $a_2$.
The contributions from non-Pomeron exchanges may be
visualised with the help of quark diagrams of Fig.1, wherein the quark structure
of Reggeons exchanged in the t-channel is seen.  
The contributions of
diagrams (1a) and (1b) differ in their phases: diagram (1a) has phase
$-\exp (-i\pi \alpha _R(t))$ with $\alpha _R(t) = 0.5 + \alpha ' t$, 
$\alpha ' \approx 1~ GeV^{-2} $ 
(for assumed SU(3) symmetry), while diagram (1b) has phase $-1$, i.e.
is purely real, in agreement with the requirement of no exotic states in the
$s$-channel. 

For Cabibbo-suppressed $D^0$ decays there are six final PP states of interest 
to us.  In the basis of definite isospin these symmetrized two-boson states are:
\begin{eqnarray}
\label{eq:defofstates}
|(\pi \pi )_2\rangle  & = & \frac{1}{\sqrt{6}}
(\pi ^+ \pi ^- + \pi ^- \pi ^+ + 2 \pi ^0 \pi ^0) \nonumber \\
|(K\bar{K})_1\rangle & =& \frac{1}{2}
(K^+ K^- + K ^- K^+ + K^0 \bar{K}^0 + \bar{K}^0 K^0) \nonumber \\
|(\pi ^0 \eta _8)_1\rangle & = & \frac{1}{\sqrt{2}}
(\pi ^0 \eta _8 + \eta _8 \pi ^0) \nonumber \\
|(\pi \pi )_0\rangle  & = & \frac{1}{\sqrt{3}}
(\pi ^+ \pi ^- + \pi ^- \pi ^+ - \pi ^0 \pi ^0) \nonumber \\
|(K \bar{K})_0 \rangle  & = & \frac{1}{2}
(K^+ K^- + K ^- K^+ - K^0 \bar{K}^0 - \bar{K}^0 K^0) \nonumber \\
|(\eta _8 \eta _8)_0 \rangle & = & \eta _8 \eta _8
\end{eqnarray}
where the subscript in $|()_I \rangle $ denotes isospin I of the state.
In terms of quark diagram amplitudes, the decays of $D^0$ to these states
are given by
\begin{eqnarray}
\label{eq:qdDdecays}
\langle (\pi \pi )_2| w | D^0 \rangle & = & \frac{1}{\sqrt{6}}
(a+b) \nonumber \\
\langle (K \bar{K})_1| w | D^0 \rangle & = & \frac{1}{2}
(a+c-e) \nonumber \\
\langle (\pi ^0 \eta _8)_1 | w | D^0 \rangle & = & \frac{1}{\sqrt{6}} 
(-b+c-e)\nonumber \\
\langle (\pi \pi )_0 | w | D^0 \rangle & = & \frac{1}{\sqrt{3}}
(a-\frac{1}{2}b+\frac{3}{2}c+\frac{3}{2}e+3f) \nonumber \\
\langle (K \bar{K})_0| w | D^0 \rangle & = & \frac{1}{2}
(a+c-e-4f) \nonumber \\
\langle (\eta _8 \eta _8)_0 | w | D^0 \rangle & = &
\frac{1}{2} (-b+c-\frac{1}{3}e-2f)
\end{eqnarray}
where quark amplitudes are denoted by $a$ (tree-level), $b$ (colour-suppressed),
$c$ (W-exchange), $e$ ("horizontal" penguin), $f$ 
(Zweig-rule violating "vertical" penguin) \cite{ChauCheng87}.

For $B^0$ decays, we have analogously
\begin{eqnarray}
\label{eq:qdBdecays}
\langle (\pi \pi )_2| w | B^0 \rangle & = & 
-\frac{1}{\sqrt{6}}(a+b) 
\nonumber \\
\langle (K \bar{K})_1| w | B^0 \rangle & = & 
\frac{1}{2}(c-e) 
\nonumber \\
\langle (\pi ^0 \eta _8)_1 | w | B^0 \rangle & = & 
\frac{1}{\sqrt{6}} (c-e)
\nonumber \\
\langle (\pi \pi )_0 | w | B^0 \rangle & = & 
-\frac{1}{\sqrt{3}}(a-\frac{1}{2}b+\frac{3}{2}c+\frac{3}{2}e+3f) 
\nonumber \\
\langle (K \bar{K})_0| w | B^0 \rangle & = & 
\frac{1}{2}(c+e+4f) 
\nonumber \\
\langle (\eta _8 \eta _8)_0 | w | B^0 \rangle & = &
\frac{1}{6} (b+c+e+6f)
\end{eqnarray}
with amplitudes $a$ (tree), $b$ (colour-suppressed), $c$ ($W$-exchange), 
etc. different from those in $D$ decays.

In Eq.(\ref{eq:qdDdecays}) we assumed SU(3) symmetry in weak
decays, i.e. equal amplitudes for the production of 
strange ($s\bar{s}$) and nonstrange quark pairs.
Below we will estimate Pomeron and non-leading Reggeon contributions both
without and with coupled-channel effects.

There are three separate non-communicating 
sectors of different isospin (I=0,1,2).   
Let us first discuss the contributions of exchange-degenerate SU(3)-symmetric
Reggeons.
The numerical factors describing the strength of various
transitions are computed by sandwiching diagrams of Fig.1 (denoted below by
$C$ for crossed diagrams and $U$ for uncrossed ones) in between 
the states of Eq.(\ref{eq:defofstates}).

In the I=2 sector one obtains (the second equality essentially defines
our normalization convention)
\begin{eqnarray}
\label{eq:I2weights}
\langle (\pi \pi)_2| U_2 |(\pi \pi)_2 \rangle & = & 0 \nonumber \\ 
\langle (\pi \pi )_2 | C_2 | (\pi \pi )_2 \rangle & = & 2
\end{eqnarray}
i.e. there is only a contribution from the crossed diagram of Fig.1b.

In the I=1 sector there are two states $(K\bar{K})_1$ and $(\pi ^0 \eta _8)_1$,
and, consequently, we have coupled-channel effects described together with
quasi-elastic effects by two 2x2 matrices.
For the uncrossed exchanges we have the matrix
\begin{equation}
\label{eq:I1weightsII}
{\bf U}_{1} = [ \langle i | {\bf U}_1 | j \rangle ] = 
\left[ \begin{array}{cc}
\epsilon ^2 & \sqrt{\frac{2}{3}} \epsilon \\
\sqrt{\frac{2}{3}} \epsilon & \frac{2}{3}
\end{array} \right]
\end{equation}
while for the crossed exchanges we have
\begin{equation}
\label{eq:I1weightsX}
{\bf C}_{1}=[ \langle i | {\bf C}_1 | j \rangle ] = 
\left[ \begin{array}{cc}
0 & -2\sqrt{\frac{2}{3}} |\epsilon | \\
-2 \sqrt{\frac{2}{3}} |\epsilon | & \frac{2}{3}
\end{array} \right]
\end{equation}
The states in the rows and columns are (from top to bottom and from 
left to right): $i,j$ = $(K\bar{K})_1$ and $(\pi ^0 \eta _8)_1 $.  
The parameter
$\epsilon $ is introduced for completeness and 
clarification: entries proportional 
to $\epsilon $ or $\epsilon ^2$ arise from the propagation of one or two 
strange (anti)quarks in the $t$-channel.  
For our SU(3)-symmetric discussion of coupled-channel effects
we shall later take $\epsilon =1$.

In the I=0 sector the resulting 3x3 matrix for uncrossed exchanges is:
\begin{equation}
\label{eq:I0weightsII}
{\bf U}_{0}=[ \langle i | {\bf U}_0 | j \rangle ] =
\left[ \begin{array}{ccc}
3 & -\sqrt{3} \epsilon & -\frac{1}{\sqrt{3}} \\
-\sqrt{3} \epsilon &2 + \epsilon ^2 & \frac{5}{3} \epsilon \\
 -\frac{1}{\sqrt{3}} & \frac{5}{3} \epsilon & \frac{1+8\epsilon ^2}{9}
\end{array} \right],
\end{equation} 
while for the crossed exchanges we have
\begin{equation}
\label{eq:I0weightsX}
{\bf C}_{0} = [ \langle i | {\bf C}_0 | j \rangle ] =
\left[ \begin{array}{ccc}
-1 & 0 & -\frac{1}{\sqrt{3}} \\
0 & 0 & -\frac{4}{3} |\epsilon | \\
- \frac{1}{\sqrt{3}} & -\frac{4}{3} |\epsilon | & \frac{1+8 |\epsilon |^2}{9}
\end{array} \right]
\end{equation}
with the rows and columns corresponding to the states
$(\pi \pi)_0$, $(K\bar{K})_0$, and $(\eta _8 \eta _8)_0$
(from top to bottom and from left to right).

The relevant $A((PP)_I)$ amplitudes are obtained by multiplying the entries of 
Eqs.(\ref{eq:I2weights}-\ref{eq:I0weightsX}) by an appropriate Regge phase and
by a factor $Rs^{\alpha _R(t)}$,
where R is the Regge residue fitted from experiment:
\begin{equation}
\label{eq:Rvalue}
R= -4 g^2(\omega ,KK) = -\frac{4}{9} g^2(\omega ,pp) = -13.1~{\rm mb}
\end{equation}
with $g^2(\omega ,pp)$ extracted from ref.\cite{BPh71}.
The residues of $\rho $, $\omega $ etc. Reggeons obtained in this way from 
Eqs.(\ref{eq:I2weights}-\ref{eq:I0weightsX}) satisfy the condition of
SU(3) symmetry.  It is known that this is a good assumption.
For the trajectories themselves SU(3) is not such a good approximation.
Nonetheless, we will accept it when estimating coupled-channel effects
since it dramatically simplifies the discussion.

Calculations of refs.\cite{GPW98} correspond to considering only diagonal
entries in the matrices of 
Eqs.(\ref{eq:I1weightsII}-\ref{eq:I0weightsX}) 
and then putting $\epsilon =0$.
When one takes into account the Pomeron contribution as well, the complete
amplitudes without the coupled-channel effects
are given (as in \cite{GPW98}) by:

\begin{eqnarray}
\label{eq:Weyerscase}
A((\pi \pi)_2) &=&i\beta_P(\pi \pi )e^{2b^{\pi}_Pt}s + 2R s^{\alpha _R(t)} 
\nonumber \\
A((\pi \pi)_0)&=& i\beta_P(\pi \pi )e^{2b^{\pi}_Pt}s + 
R(3e^{-i\pi \alpha _R(t)}-1)s^{\alpha _R(t)} \nonumber \\
A((K\bar{K})_1) &=&i\beta_P(KK)e^{2b^K_Pt}s + 
R e^{-i\pi \alpha _R(t)}s^{\alpha _R(t)} \epsilon ^2
\nonumber \\
A((K\bar{K})_0)&=& i\beta_P(KK)e^{2b^K_Pt}s +
R(2+\epsilon ^2) e^{-i\pi \alpha _R(t)}s^{\alpha _R(t)}
\end{eqnarray}

For $D$ or $B$ decays one needs the projection of Regge amplitudes on the 
$s$-channel $l=0$ partial wave.
This amounts to integrating Regge amplitudes over $t$ from $t=0$ to $-s$.
With good accuracy, the Pomeron contribution is then proportional to
$\beta _P(\pi \pi)/(2b^{\pi}_P)$ for $(\pi \pi)_{0,2}$ channels and 
$\beta _P(KK)/(2b^{K}_P)$ for $(K\bar{K})_{0,1}$ channels.
Comparison with \cite{DGPW98,GPW98} yields:
\begin{equation}
\label{eq:PomeronWeyers}
\frac{\beta_P(\pi \pi )}{2b^{\pi}_P}\cdot
\frac{2b^K_P}{\beta_P(KK)}=\\
\frac{2\beta_P(0)}{3b_P}\cdot\frac{2\tilde{b}_P}{\tilde{\beta}_P(0)}=\\
\frac{8}{3}\frac{x_{\pi \pi}}{x_{K\bar{K}}}\\ \approx 1.1 (\pm 0.3)
\end{equation}
where the two entries in between the three equality signs relate our parameters to
the corresponding parameters of ref.\cite{GPW98}.
We conclude therefore that it should be reasonable to use SU(3) symmetry for
the Pomeron contribution as well.

Using 
\begin{equation}
\label{eq:Pomeronparam}
\frac{\beta_P(\pi \pi)}{2b^{\pi}_P}=
\frac{\beta_P(KK)}{2b^K_P}=\frac{9.9~{\rm mb}}{2.75~{\rm GeV^{-2}}}=
 3.6~{\rm mb~GeV^2} = P
\end{equation}
the calculations  
of the $s$-channel $l=0$ partial waves $a((PP)_I)$
give (as in \cite{GPW98})
\begin{eqnarray}
\label{eq:Weyersrepeated}
a((\pi \pi)_2) & = & iP + \frac{2R}{\alpha '}\frac{s^{-1/2}}{\ln s} \nonumber \\
a((\pi \pi)_0) & = & iP - \frac{R}{\alpha '}
\left( 
\frac{3is^{-1/2}(\ln s + i\pi)}{\ln ^2s+\pi ^2}+
\frac{s ^{-1/2}}{\ln s}
\right)\nonumber \\
a(K\bar{K})_1)  & = & iP +\frac{R}{\alpha '}
\frac{\epsilon ^2is^{-1/2}(\ln s + i\pi )}{\ln ^2 s + \pi ^2}
\nonumber \\
a((K\bar{K})_0) & = & iP - \frac{R}{\alpha '} 
\frac{(2+\epsilon ^2)is^{-1/2}(\ln s + i\pi )}{\ln ^2 s + \pi ^2}. 
\end{eqnarray} 

In ref.\cite{GPW98} the effects of FSI on weak decay amplitudes 
are estimated through multiplying quark-level amplitudes
by hadronic phase factors, while completely 
neglecting different possible hadronic
renormalization of quark-level amplitudes of different isospin.
In the calculations of this paper we shall follow this line of reasoning to see
how coupled-channel effects {\it alone} modify the results of ref.\cite{GPW98}.
Using $s = m^2_D = 3.47~{\rm GeV^2}$ in Eq.(\ref{eq:Weyersrepeated}) one finds
the phases given in columns 3, 4 of Table 1.  It is column 3 which should
be directly compared with the results of ref.\cite{GPW98} 
(quoted in column 2 of Table 1):
columns 2 and 3 differ only by our 
simplifying assumption of SU(3) for the Pomeron amplitudes.
One can see that this assumption is reasonable:
our results are close to those of ref.\cite{GPW98}.
Column 4 includes effects of $\phi$ Reggeon exchanges (assuming SU(3))
which reduce the $|\delta ^1_K-\delta ^0_K|$
phase difference.
The phase shift in $P\bar{P}$ channel of isospin I is denoted by 
$\delta _P^I$.

TABLE 1

Comparison of calculated values of phase shifts for $D$ decays

\begin{tabular}{|l|c|c|c|c|c|c|}
\hline
phase & \multicolumn{3}{c|}{no coupled channels}
      & "experiment"
      & \multicolumn{2}{c|}{coupled channels}  \\
      & ref.\cite{GPW98} & \multicolumn{2}{c|}{Eq.(\ref{eq:Weyersrepeated}) }
      & ref.\cite{CLEO} &\multicolumn{2}{c|}{$c,e,f=0$}\\
&& $\epsilon =0$ &  $\epsilon =1 $ & &r=1 &r=-1.8\\
\hline
$\delta _{\pi }^2$  &       & $162^o$  & $162^o$ & & $162^o$ & $162^o$\\
$\delta _{\pi }^0$  &       & $ 92^o$  &  $92^o$ & & $60^o$  & $52^o$ \\
$\delta _{\pi }^2-\delta _{\pi }^0$ 
                     & $60^o\pm4^o$     & $70^o$  &   $70^o$ 
                     &$82^o \pm 10^o$   & $102^o$ &   $110^o$ \\
$\delta _{K}^1$     &       &   $90^o$ & $114^o$ & & $111^o$ & $78^o$ \\
$\delta _{K}^0$     &       &  $127^o$ & $138^o$ & & $111^o$ & $78^o$ \\
$\delta _{K}^1-\delta _{K}^0$ 
                     & $-29^o \pm 4^o$  &  $-37^o$&   $-24^o$ 
                     &$\pm (24^o \pm 13^o)$ & $0^o$ & $0^o$ \\
\hline
\end{tabular}
\\

Let us now discuss the coupled-channel effects.  We will assume from now on that
$\epsilon =1$. 
In a sector of given isospin I, the matrices ${\bf U}_{I}$ and ${\bf C}_{I}$ 
commute.  
Consequently, we may
diagonalize them simultaneously.

In the $I=0$ sector the  eigenvectors of ${\bf U}_0$ and  ${\bf C}_0$ are
\begin{eqnarray}
\label{eq:I0eigenvectors}
| {\bf 1},0 \rangle & = & \frac{1}{2\sqrt{2}}
(-\sqrt{3}|(\pi \pi)_0 \rangle + 2 |(K\bar{K})_0\rangle + 
|(\eta _8 \eta _8)_0 \rangle) \nonumber \\
| {\bf 8},0 \rangle & = & \frac{1}{\sqrt{5}}
(\sqrt{3} |(\pi \pi)_0 \rangle + |(K\bar{K})_0\rangle + 
|(\eta _8 \eta _8)_0 \rangle)\nonumber \\
| {\bf 27},0 \rangle & = & \frac{1}{\sqrt{10}}
( \frac{1}{2}|(\pi \pi)_0 \rangle + \sqrt{3} |(K\bar{K})_0\rangle - 
\frac{3\sqrt{3}}{2}|(\eta _8 \eta _8)_0 \rangle)
\end{eqnarray} 
with eigenvectors labelled by the relevant SU(3) representation.

In the $I=1$ sector the eigenvectors  of ${\bf U}_1$ and  ${\bf C}_1$ are
\begin{eqnarray}
\label{eq:I1eigenvectors}
| {\bf 8},1  \rangle & = & \frac{1}{\sqrt{5}}
(\sqrt{3} |(K\bar{K})_1\rangle + \sqrt{2} |(\pi ^0 \eta _8)_1 \rangle ) 
\nonumber \\
| {\bf 27},1 \rangle & = & \frac{1}{\sqrt{5}}
(-\sqrt{2}(K\bar{K})_1+\sqrt{3}(\pi ^0 \eta _8)_1).
\end{eqnarray}

In the $I=2$ sector there is only one state, the $|{\bf 27},2\rangle =
|(\pi \pi)_2\rangle  $.

The eigenvalues corresponding to these eigenvectors
are
\begin{eqnarray}
\label{eq:eigenvalues}
|{\bf 1},I=0\rangle \rightarrow 
& \lambda _U = \frac{16}{3} & \lambda _C = -\frac{2}{3}
\nonumber \\
|{\bf 8},I=0,1\rangle \rightarrow
& \lambda _U = \frac{5}{3}  & \lambda _C = -\frac{4}{3}
\nonumber \\
|{\bf 27},I=0,1,2\rangle \rightarrow
& \lambda _U = 0            & \lambda _C = 2.
\end{eqnarray}

Amplitudes $a({\bf 27}, I)$ in the $I=0,1,2$ sectors are all equal.
Similarly, amplitudes $a({\bf 8}, I)$ in $I=0,1$ sectors are equal.
Thus, one obtains the following three different FSI amplitudes
\begin{eqnarray}
\label{eq:SU3amplitudes}
a({\bf 1})  & = & iP - \frac{R}{\alpha '}
\left( \frac{16}{3}\frac{is^{-1/2}(\ln s + i\pi)}{\ln ^2 s + \pi ^2}+
\frac{2}{3}\frac{s^{-1/2}}{\ln s}\right) \nonumber \\
a({\bf 8})  & = & iP - \frac{R}{\alpha '}
\left( \frac{5}{3}\frac{is^{-1/2}(\ln s + i\pi)}{\ln ^2 s + \pi ^2}+
\frac{4}{3}\frac{s^{-1/2}}{\ln s}\right) \nonumber \\
a({\bf 27}) & = & iP + \frac{2R}{\alpha '} \frac{s^{-1/2}}{\ln s}.
\end{eqnarray}
The amplitude $a({\bf 27})$ must be of course equal to $a((\pi \pi)_2)$ in 
Eq.(\ref{eq:Weyersrepeated})

Due to coupled-channel effects, the observed FSI-corrected
weak decay amplitudes
$\langle (\pi \pi)_0|W|D^0\rangle  $, 
$\langle (K\bar{K})_0|W|D^0\rangle $, and
$\langle (\eta _8 \eta _8)_0|W|D^0\rangle $
become linear combinations of appropriate short-distance
quark-level amplitudes in Eq.(\ref{eq:qdDdecays}), i.e.:
\begin{equation}
\label{eq:isospinamplwithFSI}  
\langle (PP)_I|W|D^0\rangle = \sum _{\bf R} \langle (PP)_I|{\bf R},I \rangle
\langle {\bf R},I|S_{FSI}|{\bf R},I\rangle \langle {\bf R},I|w|D^0 \rangle
\end{equation}
with $\langle(PP)_I|{\bf R},I\rangle$ given in 
Eqs.(\ref{eq:I0eigenvectors},\ref{eq:I1eigenvectors}),
$\langle {\bf R},I|S_{FSI}|{\bf R},I\rangle $ describing SU(3)-symmetric
final state interactions in the $|{\bf R},I\rangle $ state,
and $\langle {\bf R},I|w|D^0 \rangle $ determined from
Eqs.(\ref{eq:qdDdecays},\ref{eq:I0eigenvectors},\ref{eq:I1eigenvectors}).

In line with the spirit of ref.\cite{GPW98} and as a simple example
we assume now that the FSI-corrected weak decay amplitudes
$\langle {\bf R}, I| W | D^0 \rangle =
\langle{\bf R},I|S_{FSI}|{\bf R},I\rangle \cdot \langle{\bf R},I|w|D^0\rangle$ 
differ from quark-level expressions 
($\langle{\bf R},I|w|D^0\rangle$) by hadronic phase factors
$\exp(i{\delta ({\bf R})})$ only, i.e. that the possible hadron-level
renormalization
of the {\em magnitude} of short-distance amplitudes is negligible.
Although in general this assumption need not be true, its violation
will not change the qualitative conclusions of this paper.

Within the considered model the phases are 
determined from Eq.(\ref{eq:SU3amplitudes}).
At $s=m^2_D=3.47~{\rm GeV^2}$ one gets 
$\delta ({\bf 1})=130 ^o $, $\delta ({\bf 8})=49^o$, and
$\delta ({\bf 27}) = 162^o$.
Even with the simplifying assumption of no FSI-induced change of magnitudes
of short-distance amplitudes,
in order to estimate how the quark-level amplitudes $a$, $b$, $c$ etc.
add up, one has to make
additional assumptions about their actual size.  
Usually, one assumes that the
factorization amplitudes $a$ and $b$ dominate, thus
neglecting the contribution from
diagrams $c$, $e$, and $f$. 
Below we shall discuss two most often considered
cases: 1) "bare" quark-level relation $a=3b$ and 
2) QCD-corrected relation $a=3rb$ with $r\approx -1.8$ 
\cite{ChauCheng87,Sorensen,KC}.
Then, all amplitudes are given in terms of a single parameter $a$, the size
of which is irrelevant when determining phase shifts.
We see from Table 1
that the SU(3)-symmetric treatment of coupled-channel effects leads
to vanishing $\delta _{K}^1-\delta _{K}^0 $ phase-shift
difference.  This is in fact true for any $a$, $b$, $c$,
provided $e$ and $f$ vanish (see also ref.\cite{Zen97}).
Comparing appropriate columns in Table 1 we see that the inclusion of 
coupled-channel
effects dramatically changes hadronic phase-shift differences in the considered
model: $\delta _{\pi }^2 - \delta _{\pi }^0 =70^o \rightarrow \approx 110^o$, 
$\delta _{K}^1 - \delta _{K}^0 \approx -30^o \rightarrow 0^o$.

One may ask if similar strong dependence on coupled-channel effects
could occur in B-decays.  
For the sake of comparison, we need
the phase shifts calculated without 
coupled-channel effects.  These phase shifts are given on 
the left-hand side of Table 2. 

TABLE 2

Comparison of calculated values of phase shifts for $B$ decays

\begin{tabular}{|l|c|c|c|c|c|c|}
\hline
phase &\multicolumn{3}{c|}{no coupled channels}  
      &\multicolumn{3}{c|}{coupled channels   } \\
      & ref.\cite{GPW98} 
      &\multicolumn{2}{c|}{Eq.(\ref{eq:Weyersrepeated})} 
      &   &\multicolumn{2}{c|}{$b=a/(3r)$, $r=-3$}\\
&& $\epsilon =0$ &  $\epsilon =1 $ & $e \gg a,b$ & $e=0.2 a$ & $e = 0.04 a$ \\
\hline
$\delta _{\pi }^2$     &           &   $112^o$   & $112^o$ 
                                   &   $112^o$   & $112^o$   &  $112^o$     \\
$\delta _{\pi }^0$     &           &    $94^o$   &  $94^o$ 
                                   &    $98^o$   &  $94^o$   &  $93^o$      \\
$\delta _{\pi }^2-\delta _{\pi }^0$ 
                 & $+11^o\pm2^o$   &    $18^o$   &  $18^o$ 
                                   &    $14^o$   &  $18^o$   &  $19^o$      \\
$\delta _{K}^1$        &           &    $90^o$   &  $83^o$  
                                   &    $85^o$   &  $85^o$   &  $85^o$      \\
$\delta _{K}^0$        &           &   $100^o$   & $103^o$   
                                   &   $110^o$   & $137^o$   &  $168^o$     \\
$\delta _{K}^1-\delta _{K}^0$ 
                 & $-7^o \pm 1^o$  &   $-10^o$   & $-20^o$   
                                   &   $-25^o$   & $-52^o$   &  $-83^o$     \\
\hline
\end{tabular}
\\

If the coupled-channel effects are taken into account, at $s=m^2_B$ 
we obtain from Eq.(\ref{eq:SU3amplitudes})
the following phase shifts for different
SU(3) channels:
$\delta ({\bf 1}) = 104^o $, 
$\delta ({\bf 8}) = 85^o$, 
$\delta ({\bf 27})= 112^o$.

In order to estimate how the considered 
coupled-channel effects affect phase-shift values  
given in the left-hand side of Table 2, 
we must again make some assumptions about the relative size
of short-distance amplitudes $a$, $b$, etc. at $s=m^2_B$.  
One expects that the dominant contribution is provided by the $a$ amplitude 
and that the amplitudes $b$ and $e$ constitute a 10-20\%
correction \cite{GHLR95}.  
Contributions from other amplitudes are expected to be
much smaller \cite{GHLR95}.
Using  $ r = (c_1+c_2/3)/(3c_2+c_1) \approx -3$, 
assuming $e/a$ in the range of $0.04-0.20$
(as estimated in \cite{G93}),
and neglecting other contributions in Eq.(\ref{eq:qdBdecays}),
we can estimate the FSI phases in B-decays.
Using the procedure of Eq.(\ref{eq:isospinamplwithFSI})
for $B$ decays
one then obtains the numbers given in the right-hand side of Table 2.

We see from Table 2 that the phases in $(\pi \pi)_I$ channels do not depend
very strongly on the inclusion of coupled-channel effects. This is also the
case for the $(K\bar{K})_1$ phase.  In fact, the latter phase
is always equal to $\delta ({\bf 8})$ because the 
$|{\bf 27},1\rangle$ state is not produced in the weak $B^0$ decay
(see Eqs.(\ref{eq:qdBdecays},\ref{eq:I1eigenvectors})).
However, the $(K\bar{K})_0$ phase changes dramatically when coupled-channel 
effects are included.
Only for $a,b\ll e$ the phase change induced by coupled-channel effects is
small.
However, if $e$ is small the relevant phase change is large.
The origin of this effect can be understood from Eqs.(\ref{eq:qdBdecays}):
due to coupled-channel effects, the $(\pi \pi)_0$ state created in a 
short-distance decay process may be converted into a $(K\bar{K})_0$ final state.
Thus, the observed final $(K\bar{K})_0$ state receives contribution
from both the short-distance $B^0\rightarrow (K\bar{K})_0$ decay
(characterized by small amplitude $e$) with $(K\bar{K})_0$ elastically
scattered into the final $(K\bar{K})_0$ state,
as well as
from the short-distance $B^0\rightarrow (\pi \pi)_0$ decay
(characterized by large amplitude $a$)
which contributes through coupled-channel rescattering effects
into the final $(K\bar{K})_0$ state.
The net result is interference of the small $e$ amplitude with some
admixture coming from the large $a$ amplitude.
If the relative absolute size of the two contributions is comparable,
one can obtain a wide range of phases for their sum.
This lies at the origin of large phases 
in the $(K\bar{K})_0$ channel of $B^0$ decays
(Table 2).
Thus, in spite of a relative weakness (as compared to elastic rescattering) 
of strangeness-exchanging Reggeon contribution 
(which induces $(\pi \pi)_0 \rightarrow (K\bar{K})_0$)
at $s=m^2_B$ , 
the generated FSI phase may be large. 

The above mechanism of generating large FSI phases in 
$B^0\rightarrow (K\bar{K})_0$ decays must also work
when the two simplifying assumptions of 1) SU(3) symmetry, and 
2) no renormalization of amplitude magnitudes, are somewhat broken.  
With SU(3) symmetry broken, one would 
expect results somewhere in between those on the left- 
and right-hand sides of Table 2.
In fact, the mechanism under discussion
should be operative, provided there is a small FSI transition
$(\pi \pi)_0\rightarrow (K\bar{K})_0$, since, 
as shown in Fig.1c, such a transition generates an effective large-distance
penguin amplitude interfering with the original short-distance penguin.
Of course, intermediate states other than $\pi \pi$ 
may also contribute in a similar way to the final $(K\bar{K})_0$ state.
Furthermore, the use of Regge amplitudes for $l=0$ may be also questioned.
All this cannot change, however, our general expectation that in explicit models
for $B^0\rightarrow (K\bar{K})_0$ decays 
it is quite likely to obtain large long-distance-induced phases.

{\em Acknowledgements}
\\
I would like to thank L. Le\'sniak for discussions and reading of the
manuscript prior to publication.

%---------------------------------------------------------------

\end{document}